\documentclass[11pt,a4paper]{article}
\usepackage{jcappub}

\usepackage{graphicx}
\usepackage{dcolumn}
\usepackage{bm}

\newcommand{\rem}[1]{ } 
\newcommand{\beq}{\begin{equation}}
\newcommand{\eeq}{\end{equation}}

\newcommand{\jcap}{{J. Cosmol. Astropart. Phys.}}
\newcommand{\mnras}{{Mon. Not. R. Astron. Soc.}}
\newcommand{\aap}{{Astron. \& Astroph.}}
\newcommand{\apjl}{{Astrophys. J. Lett.}}
\newcommand{\prd}{{Phys. Rev. D}}
\newcommand{\prl}{{Phys. Rev. Lett.}}
\newcommand{\nat}{{Nature}}

\begin{document}


\title{Plasma Constraints on the Millicharged Dark Matter}

\author{Mikhail V. Medvedev$^{1-5}$} \author{Abraham Loeb$^{1}$} 
\affiliation{$^{1}$Department of Astronomy, Harvard University, Cambridge, MA 02138}
\affiliation{$^{2}$Institute for Advanced Study, School for Natural Sciences, Princeton, NJ 08540}
\affiliation{$^{3}$Department of Astrophysical Sciences, Princeton University, Princeton, NJ 08544}
\affiliation{$^{4}$Department of Physics and Astronomy, University of Kansas, Lawrence, KS 66045}
\affiliation{$^{5}$Laboratory for Nuclear Science, Massachusetts Institute of Technology, Cambridge, MA 02139}



\emailAdd{medvedev@ku.edu}
\emailAdd{aloeb@cfa.harvard.edu}

\abstract{Dark matter particles were suggested to have an electric charge smaller than the elementary charge unit $e$. The behavior of such a medium is similar to a collisionless plasma. In this paper, we set new stringent constraints on the charge and mass of the millicharged dark matter particle based on observational data on the Bullet X-ray Cluster. }



\maketitle

\section{Introduction}

\begin{figure}
\includegraphics[scale=1.5]{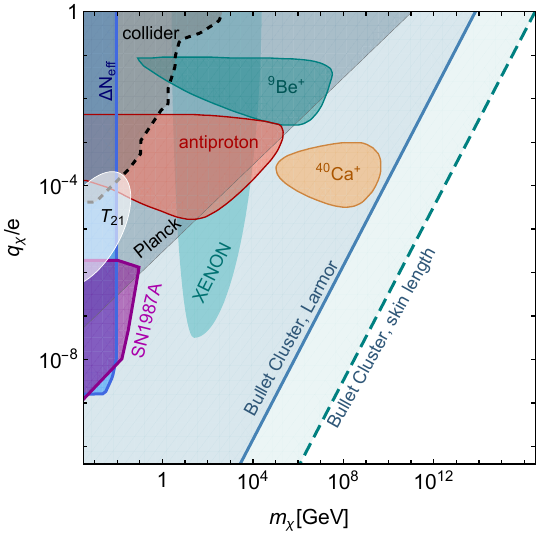}
\caption{\label{fig-excl} The charge-mass exclusion (shaded) regions for millicharged dark matter. For existing constraints, see text. The two inclined lines (solid and dashed) represent the galaxy cluster constraints discussed in this paper, Eqs. (\ref{1}) and (\ref{2}), respectively. }
\label{fig:excl}
\end{figure}

Millicharged particles with the electric charge being smaller than the unit (electron) charge, i.e., $q_\chi=\epsilon e$ with $\epsilon<1$, is a possible theoretical extension of the standard model of particle physics. Hereafter, we denote the charge magnitude by $q_\chi$, whereas individual particles can have both positive and negative charges to satisfy global charge neutrality of the medium. This possibility has been discussed extensively recently and there are numerous searches that constrain the charge-to-mass ratio of such particles, see e.g., Refs. \cite{millicharged15, 21cmConstraint, millicharged22}. The compilation of the exclusion regions is presented in Fig. \ref{fig-excl}. The regions based on experimental data are labeled as follows. ``$\rm ^{9}Be^+$'' labels the constraint from the Paul trap experiment \cite{9Be}, ``$\rm ^{40}Ca^+$'' refers to the Penning trap experiment \cite{40Ca}, ``antiproton'' refers to the cryogenic Penning trap \cite{antiproton}, ``collider'' represents the combined limits set by collider experiments, e.g., SLAC and LEP \cite{slac,lep}, ``Xenon'' denotes the limit from Xenon100 experiment \cite{xenon}. 

The 21-cm temperature anomaly reported by the EDGES Collaboration \cite{edges} received much attention and has been interpreted as a result of the interaction of baryons with millicharged dark matter \cite{Barkana18, Abo+21}. The allowed region is marked in white in Fig. \ref{fig-excl} and labeled ``$T_{21}$''. Such an interpretation has been ruled out based on astrophysical observations. The constraints include the limit on the dark-matter-baryon interaction using the CMB Planck15 data labeled ``Planck'' \cite{planck}, the implications from the light element abundances produced by Big Bang nucleosynthesis, labeled ``$\Delta N_{\rm eff}$'' \cite{Neff}, and the limit derived from observations of the nearby 1987A supernova, labeled ``SN1987A'' \cite{sn1987a}.

Despite numerous existing constraints, the possibility that the dark matter can be composed of millicharged particles is not ruled out as of yet. In this paper, we present new stringent cosmological constraints based on observations of the Bullet X-ray Cluster.

\section{Bullet Cluster constraints}

The Bullet X-ray Cluster 1E~0657-56 consists of two colliding sub-clusters and provides a strong indication for the existence of dark matter which is not coupled to the baryons. Indeed, the positions of the clumps of two collisionless components --- dark matter and stars --- do not coincide with the position of the collisional hot intracluster gas \cite{bullet1, bullet2, bullet-dm,lensing}. The separation of the two mass centroids is about $d_{c}\sim 0.3$~Mpc. The two sub-clusters collide with a relative velocity in the range $v\sim3000-4500$~km~s$^{-1}$ \cite{bullet-v, 4500km/s}. We use the value of $v\sim4000$~km~s$^{-1}$ in our estimates. It is well known that galaxy clusters possess rather strong, sub-equipartition magnetic fields of a microgauss strength. For instance, Abel 2345 has B-field in the range of $B\simeq0.3-2.8~\mu\textrm{G}$ \cite{abel2345}, which is a factor of three smaller than the equipartition value. The Bullet Cluster radio emission indicates the presence of a strong magnetic field, though its strength is poorly constrained \cite{bullet-radio14, bullet-radio23}. For our estimates, it is reasonable to assume that $B\sim 1~\mu\textrm{G}$. This is a factor of a few below the equipartition value \cite{bullet-b}. Furthermore, the very low fraction of the polarized emission --- of the order of a percent --- indicates the absence of a large scale ordered magnetic field. Instead, this is indicative of the turbulent nature of the field. 

Now, suppose that dark matter particles, denoted by $\chi$, have a small charge, $q_{\chi}$. We do not set an {\it a priori} constraint on the value of the charge, but instead use the observed properties of the Bullet Cluster to draw two constrains on the dark matter electric charge.

First, charged dark matter particles experience the Lorentz force in the intracluster magnetic fields. The characteristic Larmor radius $v/\omega_{B}$ (for a cyclotron frequency $\omega_B$) of these particles is given by
\begin{align}
r_{L}&\simeq\frac{v m_{\chi}c}{q_{\chi} B}\nonumber\\
&\simeq (4\times10^{10}\,\textrm{cm}) \left({q_{\chi}}/{e}\right)^{-1}m_{\chi,{\rm GeV}}\ v_{4000}\ B_{\mu\rm G}^{-1},
\end{align}
where the velocity $v$ is normalized to 4000~km/s, i.e., $v_{4000}=v/(4000\,{\rm km~s^{-1}})$, the particle mass is in GeV and the magnetic field is in microgauss. Here we assumed an isotropic particle distribution, so that we can approximately write ${\bf v\times B}\simeq v B$. In a turbulent B-field of a colliding and merging cluster, the Larmor scale is a characteristic distance the charged dark matter can travel ballistically, before their motion is substantially deflected by the Lorentz force. Thus, the distance between the mass centroids must be substantially smaller than the Larmor radius of dark matter particles, $d_{c}<r_{L}$. This condition yields the first charge-mass constraint:
\begin{equation}
q_{\chi}/e\lesssim 5\times10^{-14}\ m_{\chi,{\rm GeV}}\ d_{c,0.3{\rm Mpc}}^{-1}\ v_{4000}\ B_{\mu\rm G}^{-1}.
\label{1}
\end{equation}
This inequality represents an upper bound on the dark matter millicharge. It is shown in Fig. \ref{fig:excl} and labeled as ``Bullet Cluster, Larmor''. The shaded region above the line is ruled out. 
 
Second, millicharged dark matter forms a collisionless plasma throughout the universe, with a small charge-to-mass ratio. When plasma blobs collide, e.g., as in the Bullet Cluster, their plasma is susceptible to various instabilities, both kinetic and magnetohydrodynamic. The fastest ones (and most important here) are the kinetic instabilities, such as the Weibel, Buneman and two-stream instabilities. Given that kinetic energy associated with dark matter greatly exceeds that of other constituents (e.g., magnetized baryonic gas), these instabilities should be robustly excited through dark matter interactions. The kinetic plasma instabilities typically operate on the so called ``{\it plasma time scale}'', $\tau_{p,\chi}\simeq\omega_{p,\chi}^{-1}$ where the plasma frequency of species $\chi$ is
\begin{align}
\omega_{p,\chi}&=\left({4\pi q_{\chi}^{2} n_{\chi}}/{m_{\chi}}\right)^{1/2}\nonumber\\
&\simeq (3\times10^{2}\,\textrm{s})\left({q_{\chi}}/{e}\right)m_{\chi,{\rm GeV}}^{-1}\,\delta_{c,4}^{1/2},
\end{align}
where $n_{\chi}=\delta_{c}\, \rho_{\rm crit}/m_{\chi}$ is the intra-cluster millicharged dark matter density, $\rho_{\rm crit}\simeq 10^{-29}~\textrm{g~cm}^{-3}$ is the critical density of the universe, $\delta_{c,4}=\delta_{c}/10^{4}$, and $\delta_{c}$ is the cluster overdensity, characterizing how much the density of dark matter within the cluster exceeds the critical density. The value of $\delta_{c}\sim10^{4}$ is a lower limit on the density of interest in the core of the Bullet Cluster. The estimate for the value of $\delta_{c}$ follows from  the gravitational lensing map of the Bullet Cluster \citep{lensing}, which yields the dark matter density of about $2\times10^{-25}\ \textrm{g\ cm}^{-3}$ within about $0.3$~Mpc of its center, which corresponds to $\delta_{c}\sim 2\times10^{4}$.

These instabilities result in density and field inhomogeneities on kinetic scales. The Weibel instability produces filaments and associated magnetic fields elongated along the plasma motion. In contrast, the Buneman and two-stream instabilities cause plasma bunching (mediated by electric fields) into planar fronts perpendicular to the plasma motion. To understand how these processes set in and operate in the colliding cluster environment would require dedicated numerical simulations. A simple but robust estimate can, however, be made as follows. 

The general outcome of the kinetic instabilities driven by cluster collision is the formation of density fluctuations and inhomogeneities of the millicharged dark matter in the region of halo interaction. The characteristic scale of these inhomogeneities is the so-called ``{\it plasma skin length}'', 
\begin{align}
\lambda_{\chi}&\simeq c/\omega_{p,\chi}\nonumber\\
&\simeq (10^{8}\,\textrm{cm})\left({q_{\chi}}/{e}\right)^{-1}m_{\chi,{\rm GeV}}\ \delta_{c,4}^{-1/2}.
\end{align}

From an observational perspective, the absence of large dark matter inhomogeneities in gravitational lensing maps on scales of order the mass centroid separation implies that the plasma skin length is comparable or exceeds this separation, namely $\lambda_{\chi}\gtrsim d_{c}$. This condition sets the second charge-mass constraint:
\begin{equation}
q_{\chi}/e\lesssim 10^{-16}\ m_{\chi,{\rm GeV}}\ d_{c,0.3{\rm Mpc}}^{-1}\ \delta_{c,4}^{-1/2}.
\label{2}
\end{equation}
This constraint is depicted in Fig. \ref{fig:excl} by the dashed line, labeled as ``Bullet Cluster, skin length''. The shaded region above (and to the left of) the line is ruled out. Note that this second constraint is independent of the magnetic field, but depends on the dark matter over-density parameter instead.

\section{Discussion}

We derived two robust constraints on millicharged dark matter properties based on observations of the Bullet X-ray Cluster. The first constraint is set by the fact that the mass density centroid separation cannot exceed the Larmor radius of the millicharged dark matter particles. The second constraint follows from that fact that dark matter density inhomogeneities could form via collisionless kinetic instabilities excited by the collision of the two sub-clusters. The absence of substantial inhomogeneities in gravitational lensing maps on scales smaller than the mass centroid separation puts the lower limit on the millicharged plasma skin length. These constraints are given by Eqs. (\ref{1}) and (\ref{2}), and shown in Fig. \ref{fig:excl}. Interestingly, despite the different physics involved, both constraints appear very similar in that they constrain the charge-to-mass ratio, $q_{\chi}/m_{\chi}$. 

Although this study is motivated by the millicharged dark matter model in which the charge is smaller than the electron charge ($q_{\chi}<e$), the constraints can be used for {\em any} set of particles with a charge  $q_{\chi}$. For example, the constraints obtained here allow for the dark matter to be multiply charged, $q_{\chi}>e$, provided the particles are very massive, $m_{\chi}>10^{16}~\textrm{GeV}$. These include charged primordial black holes. For a typical mass of $m_{\rm bh}\sim 10^{17}~{\rm g}\simeq 10^{41}~\textrm{GeV}$, the charge is constrained to be below $q_{\rm bh}\lesssim10^{25}e$. Apparently, this is not a strong constraint on the primordial black hole charge.

The charge attained by a black hole, immersed in the hot ionized gas medium, can be estimated as follows. First, a black hole in the membrane paradigm can be considered as a conducting sphere. Second, a conductor immersed in an ionized gas of temperature $T$ attains a floating potential, $V\simeq k_{\rm B} T/e\sim10^{-7}T$, where the temperature is in Kelvin and the potential is in Volts. A sphere of charge $Q$ and radius $r$ produces an electric potential $V=Q/r=k_{\rm B} T/e$. The black hole's Schwarzschild radius is $r_{\rm bh}=2Gm_{\rm bh}/c^{2}\sim 10^{-28}m_{\rm bh}$ in cgs units. For $m_{\rm bh}\sim 10^{17}$~g, its radius is a hundred proton radii, $r_{\rm bh}\sim10^{-11}~\textrm{cm}$. Therefore, we estimate the charge as
\begin{align}
q_{\rm bh}\sim\frac{2Gk_{\rm B}}{ec^{2}}\,T\,m_{\rm bh}
\sim(0.8e
)\,T_{8.3}\,m_{{\rm bh},17},
\end{align}
where in the second equality the charge is expressed in the units of an elementary charge, $m_{\rm bh}$ is normalized to $10^{17}$~g, and $T$ is normalized to $2\times10^{8}$~K  (i.e., $T_{8.3}=T/10^{8.3}\,{\rm K}$) which  corresponds to the Bullet Cluster gas temperature of about 17.4~keV \citep{17.4keV}. Thus, we see that a typical charge of a primordial black hole immersed in a hot ionized ambient gas is many orders of magnitude smaller than the limit set by the Bullet Cluster data.


\bibliography{millichDM}

\end{document}